\documentclass{PoS}

\usepackage{amsmath}
\usepackage{graphicx}
\usepackage{wrapfig}

\newcommand{\gev}{{\,{\rm GeV}}}
\newcommand{\Tsub}{\widetilde{T}_1}

\title{Scaling and higher twist in the nucleon Compton amplitude}

\ShortTitle{Scaling and higher twist in the nucleon Compton amplitude}

\author{A.~Hannaford-Gunn,$^a$
  R.~Horsley,$^b$
  Y.~Nakamura,$^c$
  H.~Perlt,$^d$
  P.~E.~L.~Rakow,$^e$
  G.~Schierholz,$^f$
  K.~Somfleth,$^a$
  H.~St\"uben,$^g$
  \speaker{R.~D.~Young},$^a$
  J.~M.~Zanotti,$^a$\\
  \\{\bf QCDSF-UKQCD-CSSM Collaboration}\\
  \\
        \llap{$^a$}CSSM, Department of Physics, University of Adelaide, Adelaide SA 5005, Australia\\
        \llap{$^b$}School of Physics and Astronomy, University of Edinburgh, 
        Edinburgh EH9 3FD, United Kingdom\\
        \llap{$^c$}RIKEN Advanced Institute from Computation Science, Kobe, Hyogo 650-0047, Japan\\
        \llap{$^d$}Institut f\"ur Theoretische Physik, Universit\"at Leipzig, 04103 Leipzig, Germany\\
        \llap{$^e$}Theoretical Physics Division, Department of Mathematical Physics,
                   University of Liverpool, Liverpool L69 3BX, United Kingdom\\
        \llap{$^f$}Deutsches Elektronen-Synchrotron DESY, 23603 Hamburg, Germany\\
        \llap{$^g$}RRZ, Univeristy of Hamburg, 20146 Hamburg, Germany\\
        E-mail: \email{ross.young@adelaide.edu.au}}

\abstract{The partonic structure of hadrons plays an important role in a vast array of high-energy and nuclear physics experiments. It also underpins the theoretical understanding of hadron structure. Recent developments in lattice QCD offer new opportunities for reliably studying partonic structure from first principles. Here we report on the use of the Feynman-Hellmann theorem to study the forward Compton amplitude in the unphysical region. We demonstrate how this amplitude provides direct constraint on hadronic inelastic structure functions. The use of external momentum transfer allows us to study the $Q^2$ evolution to explore the onset of asymptotic scaling and reveal higher-twist effects in partonic structure.}

\FullConference{The 37th Annual International Symposium on Lattice Field Theory - LATTICE2019\\
		16-22 June, 2019\\
		Wuhan, China.}

\begin{document}

\section{Introduction}
Hadron structure functions are ubiquitous in the description of leptonic interactions with hadrons, encoding: elastic form factors, inclusive electro-(and photo-)production of resonances, diffractive processes and Regge phenomena, and partonic structure in deep inelastic scattering.
Lattice QCD, however, has really only been able to probe some limited kinematic corners of the all-encompassing structure functions---primarily being limited to elastic form factors and low(est) moments of leading-twist parton distributions.
The particular interest in partonic structure has motivated a number of strategies to overcome limitations in the lattice formulation, including: the Euclidean hadron tensor \cite{Liu:1993cv}, lattice OPE \cite{Gockeler:1995wg,Gockeler:1996mu,Capitani:1998fe}, heavy-quark Compton amplitude \cite{Detmold:2005gg}, symmetry-improved operator construction \cite{Davoudi:2012ya}, factorisable matrix elements\footnote{Nomenclature attributable to Monahan \cite{Monahan:2018euv}.} \cite{Ma:2017pxb}, and the most popular quasi-PDFs \cite{Ji:2013dva,Lin:2014zya} and related quantities \cite{Orginos:2017kos}. 

We have embarked on a complementary program to extract the forward Compton amplitude in the {\em unphysical} region \cite{Chambers:2017dov}---a similar strategy was also suggested by Ji and Jung in Ref.~\cite{Ji:2001wha}.
From a theoretical perspective, this approach is similar to the Euclidean hadron tensor and heavy-quark Compton amplitude, however, respectively, we avoid making connection between Euclidean and Minkowski time coordinates and exploit physical kinematics to ensure the current-current separation remains spacelike.
Computationally, we are able to take advantage of the efficiency of the Feynman-Hellmann approach to hadron structure \cite{Horsley:2012pz,Chambers:2014qaa} and avoid the need to compute 4-point functions.
Building upon the exploratory study of Ref.~\cite{Chambers:2017dov}, here we highlight some recent progress towards revealing scaling and higher twist-phenomena in the low-order moments of the Compton amplitude from lattice QCD.
For the reconstruction of the $x$-dependent parton distributions, see Ref.~\cite{holger} in these proceedings.

\section{Compton amplitude}
In this section, we review the familiar features of the hadron tensor and Compton amplitude, and present our notation.
The general description for charged lepton scattering from a hadronic target is encoded in the hadron tensor:
\begin{align}
W_{\mu\nu}(p,q)=\frac{1}{4\pi}\int d^4x\, e^{iq\cdot x}\langle p|[J_\mu(x),J_\nu(0)]|p\rangle.
\end{align}
This can be expressed in terms of the Lorentz decomposition\footnote{This decomposition is a commonly used form (e.g.~the PDG \cite{Tanabashi:2018oca}), chosen such that the structure functions map onto the familar scaling functions in the deep-inelastic scattering region, $F_1= \tfrac12(q+\bar{q})$ and $F_2=x(q+\bar{q})$ (with quark charges set to unity). Here we will assume spin-averaged quantities, hence spin indices and averaging will be suppressed.}
\begin{align}
W_{\mu\nu}(p,q)=\left(-g_{\mu\nu}+\frac{q_\mu q_\nu}{q^2}\right)F_1(p\cdot q,Q^2)
+\frac{1}{p\cdot q}\left(p_\mu-\frac{p\cdot q}{q^2}q_\mu\right)\left(p_\nu-\frac{p\cdot q}{q^2}q_\nu\right)F_2(p\cdot q,Q^2).
\label{eq:lorentz}
\end{align}
The structure functions are related to the imaginary part of the forward Compton amplitude by the optical theorem, $F_i=\mathrm{Im}\,T_i/(2\pi)$.
The $T_i$ are the corresponding scalar functions (analogous to Eq.~(\ref{eq:lorentz}))
of the forward Compton amplitude:
\begin{align}
  T_{\mu\nu}(p,q)=i\int d^4x\, e^{iq\cdot x}\langle p|{\rm T}\left\{J_\mu(x)J_\nu(0)\right\}|p\rangle.
  \label{eq:compton}
\end{align}
%
%

The Compton amplitude $T$, at fixed $Q^2$, is an analytic function of
the variable $p\cdot q$ with discontinuities associated with inelastic
particle production for $|p\cdot q|>Q^2/2$.
For convenience, we adopt the variable $\omega=2p\cdot q/Q^2$.
Below the elastic threshold (occuring at $|\omega|=1$),
the amplitude is purely real and permits a dispersive representation
in terms of an integral along the cut:
\begin{align}
T_1(\omega,Q^2)-T_1(0,Q^2)
=\frac{4\omega^2}{2\pi}\int_1^\infty d\omega'\,\frac{\mathrm{Im}\,T_1(\omega',Q^2)}{\omega'(\omega^2-\omega'^2)}
=4\omega^2\int_0^1dx\,x\frac{F_1(x,Q^2)}{1-(\omega x)^2}.
\label{eq:T1sub}
\end{align}
At the final equality, the integral has been transformed to describe an integral over the Bjorken-$x$ variable $x=1/\omega'$.
The integral has been once subtracted, 
owing to the divergent $\omega'$ behaviour of $F_1$.
In the following, where necessary, we will the use of the shorthand
notation $\Tsub$ to denote the subtracted quantity.
%
Expanding the geometric series in Eq.~(\ref{eq:T1sub}) gives
\begin{align}
  \Tsub(\omega,Q^2)=4\sum_{j=1}^\infty t_{1,2j-1}(Q^2)\omega^{2j},
  \label{eq:expansion}
\end{align}
where the expansion coefficients are given by moments of the structure function:
\begin{align}
  t_{1,2j-1}(Q^2)=\int_0^1 dx\,x^{2j-1}F_1(x,Q^2).
  \label{eq:moments}
\end{align}
Note that there is a singularity at $|\omega|=1$ on the RHS of Eq.~(\ref{eq:T1sub}), which gives rise to a branch point in $\Tsub(\omega)$.

At sufficiently large $Q^2$, $F_1$ will be dominated by the
leading-twist partonic structure.
Here, in the parton model limit, the
moments of the structure function correspond directly to the moments
of the leading-twist parton distributions $t_{1,2j-1}^{\rm PM}=\langle
x^{2j-1}\rangle/2$.\footnote{The ${\rm PM}$ superscript denotes the parton model.}
Beyond the parton model, the evolution in $\log Q^2$ is calculable in
perturbative QCD.
A feature of the present formalism is that, in principle, one can also
study the transition to low $Q^2$---where higher-twist terms
become numerically relevent, and even into the genuinely
nonperturbative domain beyond the operator product expansion.

\section{Compton amplitude on the lattice}
The Feynman-Hellmann relation allows one to relate energy shifts in a
weak external field to matrix elements of corresponding
operators---see \cite{Chambers:2014qaa} for general presentation on
the application in lattice field theory.
The extension to second order is rather straightforward \cite{Chambers:2017dov}, the details
and lattice subtleties will be presented in a forthcoming publication
\cite{kim} (see also \cite{Agadjanov:2018yxh}).
To access the Compton amplitude, the quarks are coupled to a
spatially-varying external vector potential by the modification to the
action:
\begin{align}
  S_0\to S_\lambda=S_0+\lambda_\mu\sum_x 2\cos(\mathbf{q}\cdot
  \mathbf{x}) J_\mu(x).
\end{align}

By a straightforward application of second-order time-independent perturbation theory, with relativistic normalisation of states, the quadratic energy shift is given by:
\begin{equation}
  \left.\frac{\partial^2 E}{\partial\lambda_\mu^2}\right|_{\lambda\to 0}
  =2\sum_X\frac{1}{2E_{X(\mathbf{p}+\mathbf{q})}}\frac{\langle p|J_\mu|X(\mathbf{p}+\mathbf{q})\rangle\langle X(\mathbf{p}+\mathbf{q})|J_\mu|p\rangle}{E_p-E_{X(\mathbf{p}+\mathbf{q})}}+\left(\mathbf{q}\to -\mathbf{q}\right),
  \label{eq:sumX}
\end{equation}
where the sum runs over all intermediate states of definite momentum $\mathbf{p}\pm\mathbf{q}$.
This is precisely the same sum that appears in the direct evaluation of the time-ordered product in Eq.~(\ref{eq:compton}), such that we obtain the result:
\begin{equation}
  \left.\frac{\partial^2 E}{\partial\lambda_\mu^2}\right|_{\lambda\to 0}=-\frac{1}{E_p}T_{\mu\mu}(p,q),
\end{equation}
provided one avoids the singularity in Eq.~(\ref{eq:sumX}), such that $E_p<E_{X(\mathbf{p}\pm\mathbf{q})}$.
Physically, this restriction is just the same, as above, that one must stay below the (in)elastic threshold, $|p\cdot q|< Q^2/2$.

Building upon the first numerical results reported in Ref.~\cite{Chambers:2017dov}, we carry out an extensive study of the Compton amplitude.
Here, results are performed on a single $32^3\times 64$ ensemble at an SU(3) symmetric point ($\kappa=0.12090$) \cite{Bietenholz:2011qq} using a non-perturbatively improved clover action \cite{Cundy:2009yy}.
We restrict ourselves to the third component local vector current $J_3=\bar{q}\gamma_3 q$, with $q_3=0$ and
$p_3=0$, such that the energy shift simply isolates
$T_{33}(p,q)=T_1(p\cdot q,Q^2)$.

\begin{wraptable}{r}{55mm}
\caption{
  At fixed $\mathbf{q}$, an example of the $\omega$ values probed by changing the Fourier momentum on the
  hadron state.}
\label{tab:omega}
\begin{tabular}{ccr}
       $\mathbf{q}L/(2\pi)$ & $\mathbf{p}L/(2\pi)$ & $\omega$ \\
\hline
$(4,1,0)$ & $(1,0,0)$  & $8/17$ \\
        & $(1,1,0)$    & $10/17$ \\
        & $(1,-1,0)$   & $6/17$ \\
        & $(0,1,0)$    & $2/17$ 
\end{tabular}
\end{wraptable}
For each choice of $\mathbf{q}$, a new propagator must be caclulated in the presence of the weak external field.
Given that the field strengths are weak, the free-field solution serves as a useful starting point for the conjugate-gradient inversions.
This makes the Feynman-Hellmann propagators relatively economical to compute.
Also, for each choice of $\mathbf{q}$ a range of $\omega$ values can be accessed by changing the hadron momentum.
To highlight this, in Table~\ref{tab:omega} we give an example set of $\omega$ values probed with the momentum $\mathbf{q}=(4,1,0)2\pi/L$.
%


\section{Structure function moments}
The lattice Compton amplitude is 
analysed to determine the low moments of the structure function.
This is most readily achieved by fitting the Compton amplitude using the Taylor series representation of Eq.~(\ref{eq:expansion}).
Unlike a conventional Taylor series fit, the moments are constrained to be positive definite and monotonically decreasing:
$t_{1,1}\ge t_{1,3}\ge t_{1,5}\ge\ldots\ge 0$.
From a fitting perspective, this is advantageous since the series expansion is rapidly converging and stable to the order of truncation.
While these inequalities add complication to a standard least-squares
analysis, they are straightforward to implement in a Bayesian-style analysis.

For a simplest analysis, the leading moment $t_{1,1}$ is sampled uniformly on the range
$(0,t_{\rm max})$.
A finite set of subsequent moments are then uniformly sampled in the
range $t_{1,k+2}\in (0,t_{1,k})$.
The sequences of $t_{1,k}$ are selected according to the likelihood
$\exp(-\chi^2/2)$, using the usual definition for the (correlated)
$\chi^2$.
For the simple case of fitting a single Compton amplitude, a naive
Monte Carlo sampling of the function space is sufficient.
Of course a more efficient importance-sampling algorithm could be used
for more complicated higher-dimensional fits.

\begin{wrapfigure}{r}{0.6\textwidth}
    \includegraphics[width=0.6\textwidth]{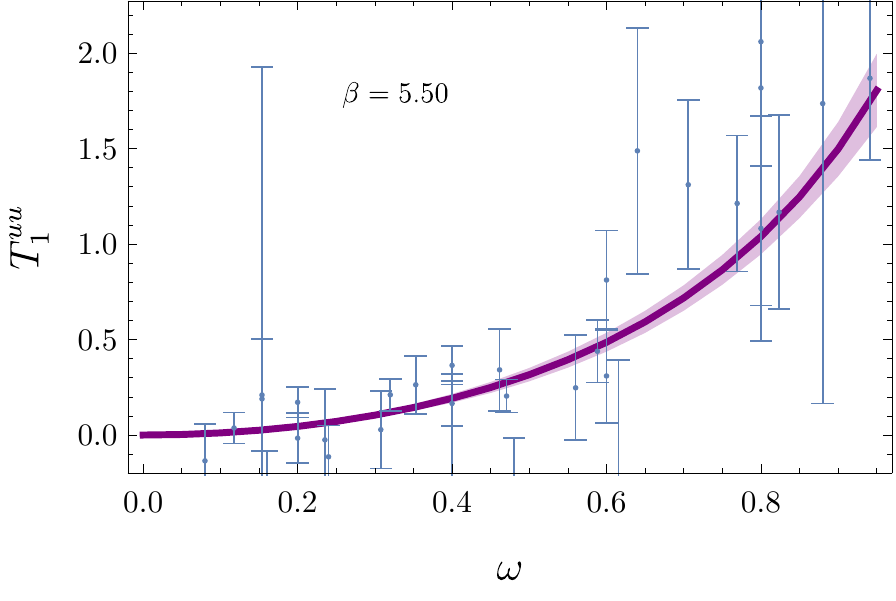}
    \vspace*{-10mm}
    \caption{Compton amplitude extracted as a function of $\omega=2p\cdot q/Q^2$ for five different $Q^2$ values ($2.7\ldots 6.9\gev^2$). The curve shows a combined fit, as described in the text, including the leading-order perturbative QCD evolution (at a factorisation scale $\mu^2=4\gev^2$).}
  \label{fig:T1combined}
\end{wrapfigure}
\begin{figure}
  \begin{center}
    \includegraphics[width=0.48\textwidth]{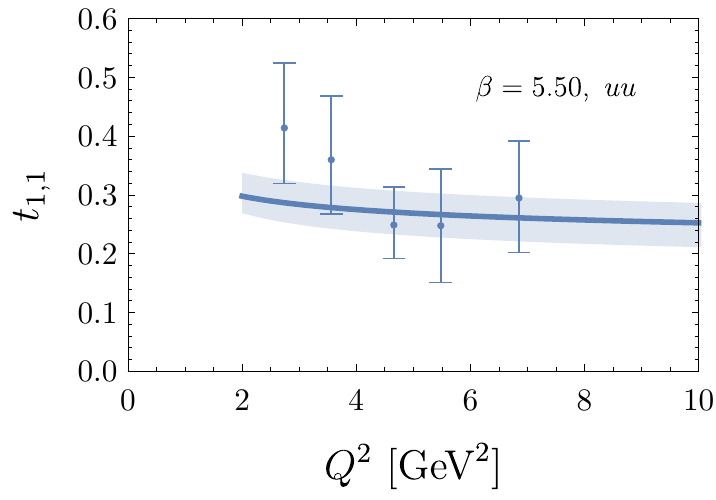}
    \includegraphics[width=0.48\textwidth]{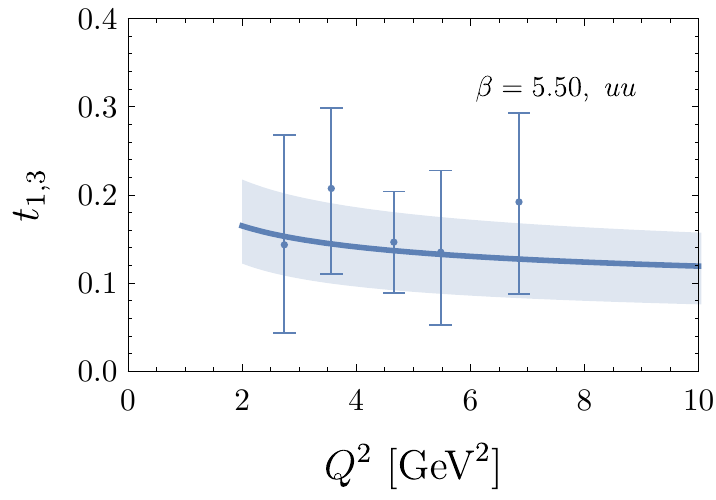}
  \end{center}
  \vspace*{-6mm}
  \caption{The data points display the extraction of the lowest two
    moments of $T_1$ from independent fits at fixed $Q^2$. The error
    bars on the points indicate the 68\% confidence level of the
    parameter determination. The curve
    displays the result of a combined fit, including the leading 
    predicted behaviour of perturbative QCD.}
  \label{fig:moments}
\end{figure}

In Figure~\ref{fig:T1combined}, we show the $u$-quark only contribution to the Compton
amplitude as a function of $\omega$ for a selection of $Q^2$ values.
We first perform a Bayesian fit, as described, to the Compton
amplitude at single values of $Q^2$ independently.
For each $Q^2$, we are able to resolve clear signals for the lowest two moments,
as displayed in Fig.~\ref{fig:moments}.
The results are certainly consistent with the Bjorken scaling behaviour
anticipated from perturbative QCD, however the statistical precision is limited.

To improve the statistical signal, we consider fitting all $Q^2$
values simultaneously.
For the combined fit, we include the leading perturbative evolution
of the moments.
The $\omega$ dependence of the combined fit is shown in
Figure~\ref{fig:T1combined} and the corresponding evolution of the
moments in Figure~\ref{fig:moments}.
We observe that the predicted $Q^2$ evolution is rather
mild,\footnote{Of course one should expect to vary $Q^2$ over orders of
magnitude in order to resolve logarithmic evolution.} and it is clear
that nothing definitively can be said about high-twist with the
present statistics.

To investigate higher twist, we look to a contribution that is expected to 
vanish in the scaling region.
In particular, we consider the interference structure function, where
one current couples to the $u$ quark and the other to the $d$.
We choose a normalisation on the interference term such that the full
proton structure function, with quark charges, is given by:
\begin{equation}
  T_1^p=\frac49 T_1^{uu}+\frac19 T_1^{dd} - \frac29 T_1^{\{ud\}}.
\end{equation}
Here, $\{ud\}$ indicates symmetrisation over the flavour indices $T^{\{ud\}}=T^{ud}+T^{du}$.

\begin{figure}
  \begin{center}
    \includegraphics[width=0.48\textwidth]{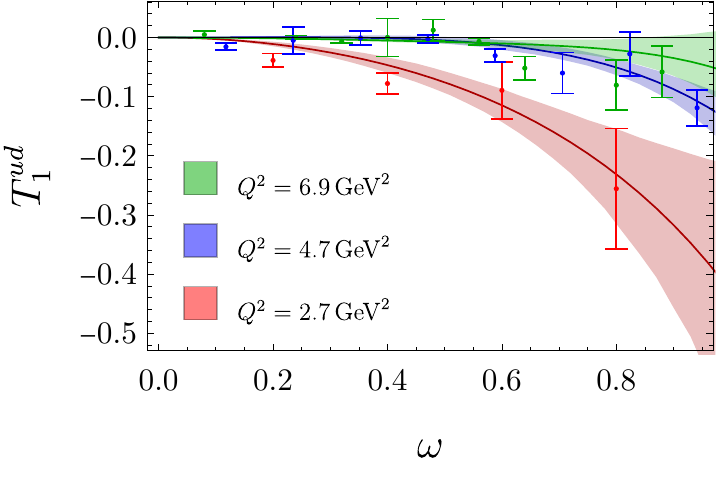}
    \includegraphics[width=0.48\textwidth]{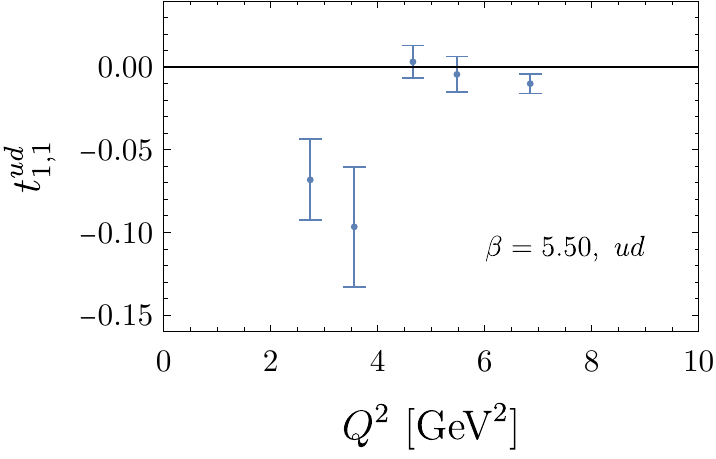}
  \end{center}
  \vspace*{-6mm}
  \caption{Compton amplitude for the $ud$ interference structure function (left panel) and the $Q^2$ dependence of the corresponding leading moment (right panel).}
  \label{fig:T1ud}
\end{figure}

The interfence term can be isolated by considering appropriate
combinations of background field strengths with differing signs on the
coupling to $u$ and $d$ quarks.
In the left panel of Figure~\ref{fig:T1ud} we show the interference structure function
for three values of $Q^2$.
In comparison to the $uu$ Compton amplitude, the interference contribution is noticably smaller in magnitude 
and there appears to be a much clearer dependence on $Q^2$.
In particular, the signal seems to be significantly suppressed at
higher-$Q^2$, in line with the expectation that this term should
vanish in the scaling region.

We repeat a similar Bayesian analysis of the $\omega$ depedence, as above, to
extract the leading moment of the interference structure function.
In this case, we don't have the same positivity bound as for the
flavour-diagonal contributions.
However, since the total structure function should be positive for any
value of the quark charges, the interference term must satisfy the constraint:
\begin{align}
\left|\widetilde{T}_1^{\{ud\}}(\omega,Q^2)\right|^2\le 4 \widetilde{T}_1^{uu}(\omega,Q^2)\widetilde{T}_1^{dd}(\omega,Q^2).
\end{align}
Similarly, since each moment can be written as an integral over a
``cross section'', an analagous inequality holds moment-by-moment
in the series expansion in $\omega$.
We sample the
interference moments uniformly within the positivity bounds dictated
by the corresponding flavour-diagonal terms.
The resultant fits are shown by the continuous curves in the left
panel of Fig.~\ref{fig:T1ud}.

The $Q^2$ dependence of the lowest moment of the $ud$ structure
function is shown in the right panel of Figure~\ref{fig:T1ud}.
Here we directly see the emerging result that the interference term is
suppressed at large $Q^2$ and a signal is apparent at low
$Q^2$---and hence has the natural interpretation as a {\em
  higher-twist} effect.
Further details on this novel observation will be reported in a
forthcoming publication.


\section{Summary}
We have presented recent progress on the study of the nucleon Compton amplitude in lattice QCD.
In particular, we have demonstrated that we can directly probe partonic structure.
Importantly, the partonic nature is accessible by kinematic selection, rather than working at the level of effective operators.
Our results show consistency with scaling phenomena that is expected at large $Q^2$, and reveal a distinct signature of higher-twist effects in the interference $ud$ structure function.

%
%

\section*{Acknowledgements}
The numerical configuration generation (using the BQCD lattice QCD
program \cite{Haar:2017ubh})) and data analysis (using the Chroma software library
\cite{Edwards:2004sx}) was carried out on the IBM BlueGene/Q and HP Tesseract using
DIRAC 2 resources (EPCC, Edinburgh, UK), the IBM BlueGene/Q (NIC,
J\"ulich, Germany) and the Cray XC40 at HLRN (The North-German
Supercomputer Alliance), the NCI National Facility in Canberra,
Australia (supported by the Australian Commonwealth Government) and
Phoenix (University of Adelaide).
RH is supported by STFC through grant ST/P000630/1.
HP is supported by DFG Grant No. PE 2792/2-1.
PELR is supported in part by the STFC under contract ST/G00062X/1.
GS is supported by DFG Grant No. SCHI 179/8-1.
RDY and JMZ are supported by the Australian Research Council grant
DP190100297.

\bibliographystyle{JHEP}
\bibliography{latt19RDY}

\end{document}